\documentstyle[preprint,aps,epsfig]{revtex}
\tighten
\newcommand{\beq}{\begin{eqnarray}}
\newcommand{\eeq}{\end{eqnarray}}

\def\acp{{\cal A}_{CP}}

\def\calb{ {\cal BR} }
\def\rmin{ (R_1)_{min} }

\newcommand{\non}{\nonumber\\ }
\title{ Extraction of the CKM angle $\gamma$ from the new "mixed " system of
$B^+ \to \pi^+ K^0$ and $B_d^0 \to \pi^0 K^0$ decays }
\author{ Zhenjun Xiao \thanks{E-mail: zjxiao@email.njnu.edu.cn}\\
Department of Physics, Nanjing Normal University, Nanjing,  Jiangsu, 210097,
People's Republic of China \thanks{Mailing address} \\
\vspace{0.5cm}
 Minping Zhang \\
Department of Physics, Henan Normal University, Xinxiang, Henan, 453002,
People's Republic of China }
\date{\today}
\begin{document}
\maketitle
\begin{abstract}
In this paper we try to extract the CKM angle $\gamma$  from the new "mixed" system
of  $B^+ \to \pi^+ K^0$ and $B_d^0 \to \pi^0 K^0$ decays. We also made an update
for the constraints on the angle $\gamma$ from the observables $R$ and $A_0$.
In the parametrization, the $SU(2)$ isospin symmetry of strong interactions
has been applied. We found the following results:
(a) the measured value of $R$ is now very close to unit, the bound on the angle $\gamma$
from the measurement of $R$ is therefore not as promising as before, but some bounds on
$\gamma$ can still be read off from $r-\gamma$ plane if $r$ could be fixed by using an
additional input;
(b) the measured $R_1$ implies a limit on the strong phase $\Delta_1$;
(c) due to the contribution from the  color allowed electroweak penguin, the minimal value
of $R_1$ can be larger than unit. For $\epsilon_1=0.2$ and $R_1=1.2$, the
range of $65^\circ \leq \gamma \leq 115^\circ$ will be excluded, such bounds on $\gamma$
are interesting and complimentary to the limits from global fit;
(d) the dependences of extraction of $\gamma$ on the variation of parameters
$\epsilon_1$, $\rho$, $r_1$ and strong phases are also studied.
\end{abstract}
%%%%%
\vspace{.5cm}

\newpage
\section{ Introduction}\label{sec-1}

As is well known, one of the main goals of the B-factories is to measure the
CKM angles $\alpha, \beta$ and $\gamma$\cite{slac504,buras01}. For the
determination of  the angle $\gamma$, $B \to \pi K, \pi \pi$ decay modes play a key
role, and have been studied intensively in the literature
\cite{prd57-2752,prd57-6843,epjc6-451,plb441,epjc11-93}. Up to now, many
two-body charmless B meson decays have been observed by
CLEO, BaBar and Belle collaborations\cite{cleo2000,babar2001,belle2001}.
For the four $B \to K\pi$ decay modes considered here, the latest world average
of the corresponding branching fractions are the following
\beq
\calb(B \to \pi^\pm K^\mp)&=&(17.3\pm 1.5)\times 10^{-6}, \non
\calb(B^\pm \to \pi^0 K^\pm )&=&(12.1 \pm 1.7)\times 10^{-6},\non
\calb(B^\pm \to \pi^\pm K^0)&=&(17.4\pm 2.6)\times 10^{-6},\non
\calb(B \to \pi^0 K^0  )&=&(10.4 \pm 1.7)\times 10^{-6},\label{eq:brkpi}
\eeq
The accuracy of the data is currently $10\%$ to $20\%$, and will be improved rapidly along
with the progress of the experiments.

For the four $B \to K \pi $ decays, the isospin and $SU(3)$ flavor symmetries of
strong interactions imply some important relations among their decay amplitudes
\cite{gronau95}. Based on these amplitude relations, three combinations of
CP-averaged $B \to \pi K$ branching ratios and the corresponding "pseudo-asymmetries"
have been considered \cite{prd57-2752,prd57-6843,epjc6-451,plb441,epjc11-93} to
probe the angle $\gamma$:
\beq
\left ( \begin{array}{l} R \\  A_0 \\ \end{array} \right ) &\equiv &
\frac{\calb (B_d^0\to  \pi^- K^+) \pm \calb (\overline{B_d^0} \to \pi^+ K^-)}{
\calb (B^+\to \pi^+ K^0) + \calb (B^- \to \pi^- \overline{K^0} )}\label{eq:rm1}\\
\left ( \begin{array}{l} R_c \\  A_0^c \\ \end{array} \right ) &\equiv &
2 \frac{\calb (B^+ \to \pi^0 K^+) \pm  \calb (B^- \to \pi^0 K^-)}{
\calb (B^+\to \pi^+ K^0) + \calb (B^- \to  \pi^-\overline{K}^0 )},
\label{eq:rc1}\\
\left ( \begin{array}{l} R_n \\  A_0^n \\ \end{array} \right ) &\equiv &
\frac{1}{2}\frac{\calb (B_d^0\to \pi^- K^+) \pm   \calb (\overline{B_d^0} \to \pi^+ K^-)}{
\calb (B_d^0 \to \pi^0 K^0) + \calb (\overline{B_d^0} \to \pi^0 \overline{K^0} )},
\label{eq:rn1}
\eeq
where the factors of $2$ and $1/2$ have been introduced to absorb the $\sqrt{2}$ factors
originating from the wavefunctions of $\pi^0$ meson. When CLEO firstly reported their
observation of the decays $B_d \to \pi^\pm K^\mp $, $B^\pm \to \pi^\pm K$,
the measured ratio $R=0.65 \pm 0.40$ lead to an interesting bound on angle
$\gamma$\cite{prd57-2752}. Since the measured $R$ is now very close to 1, however, the
constraint on the angle $\gamma$ from this "mixed" system is becoming weak now.
For the possible constraints on $\gamma$ derived from the "charged"  and "neutral" systems,
one can see for example Refs.\cite{plb441,epjc11-93} and references therein.

In this paper, we define and study a new "mixed" system $B^+ \to \pi^+ K^0$ and
$B_d^0 \to \pi^0 K^0$, to see if we can  extract out or put some constraints on the angle
$\gamma$ from the new observables, the ratio $R_1$ and the corresponding "pseudo-asymmetry"
$A_1$,
\beq
\left ( \begin{array}{l} R_1 \\  A_1 \\ \end{array} \right )
&\equiv & 2 \frac{\calb (B_d^0 \to \pi^0 K^0) \pm  \calb (\overline{B}_d^0 \to \pi^0
\overline{K}^0 )}{ \calb (B^+\to \pi^+ K^0) + \calb (B^- \to \pi^- \overline{K}^0)}
\label{eq:rs1}
\eeq
We will also make an update for the constraint on the angle $\gamma$ from the observables
$R$ and $A_0$.

Using the CP-averaged branching ratios as given in Eq.(\ref{eq:brkpi}) one finds that
\beq
R&=& 0.99\pm 0.17, \ \ R_1 = 1.20 \pm 0.36.  \label{eq:data}
\eeq
The central value of $R$ is very close to unit now.

This paper is organized as follows. In Sec.\ref{sec-2}, we present the general
description of the $B \to \pi K$ decays, define the observables and make
estimations about their magnitude. In Sec.\ref{sec-3}, we consider the new measured values
of $R$ and $A_0$ to make an update for the bounds on $\gamma$ derived from the so-called
"mixed" system: $B^+ \to \pi^+ K^0$ and $B_d^0 \to \pi^+ K^-$ decay modes.
In Sec.\ref{sec-4}, we study the new "mixed" system, $B^+ \to \pi^+ K^0$ and $B_d^0 \to \pi^0 K^0$
decays, to find the possible bounds on $\gamma$  from this new combination.
The conclusions are included in the final section.

\section{General Description of $B \to \pi K$ decays } \label{sec-2}

\begin{figure}[t]%fig.1
\vspace{-40pt}
\begin{minipage}[]{1\textwidth}
\centerline{\epsfxsize=1.2 \textwidth \epsffile{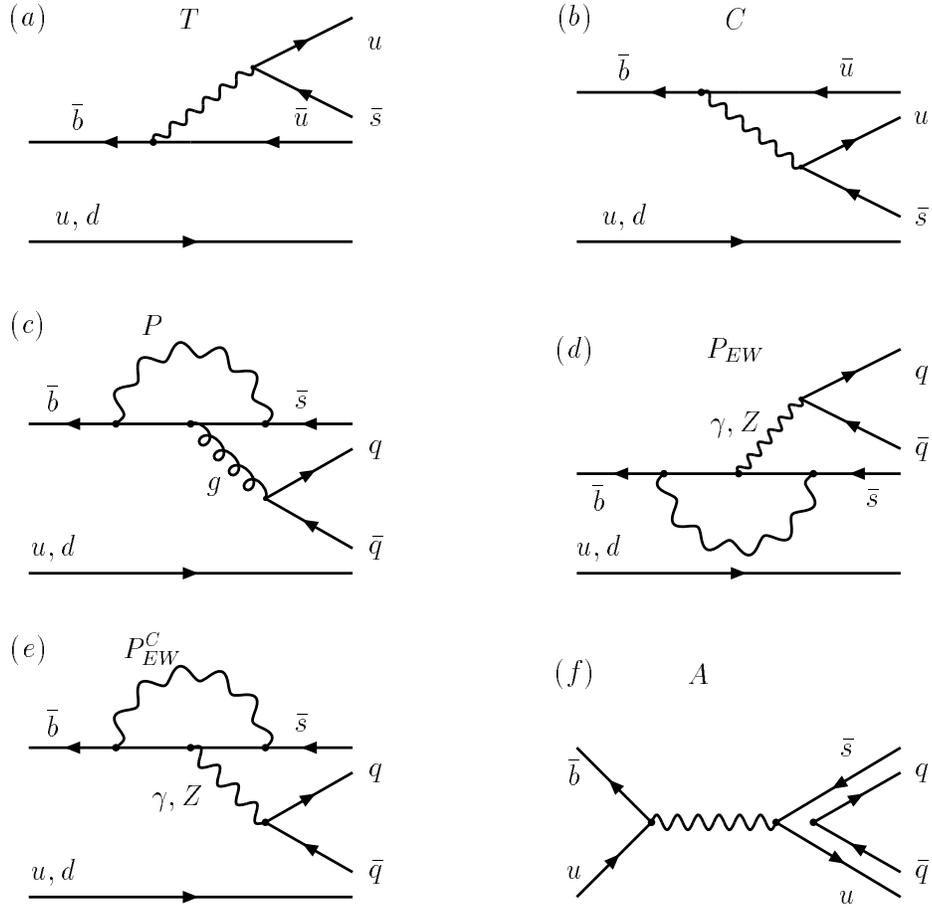}}
\vspace{-200pt}
\caption{Feynman diagrams contributing to $B \to K \pi$ decays. The $q$ denotes
the $u$ or $d$ quarks.}
\label{fig:feynman}
\end{minipage}
\end{figure}

First of all, as illustrated in Fig.\ref{fig:feynman}, the Feynman diagrams contributing
to the charmless $B \to \pi K$ decays can be classified as follows \cite{gronau95}:
\begin{itemize}
\item
a color-favored "tree" amplitude $T$ and  a color-suppressed "tree" amplitude $C$;

\item
a QCD penguin amplitude $P$;

\item
an color-allowed electroweak (EW) penguin amplitude $P_{EW}$, and a color-suppressed EW
penguin amplitude $P^C_{EW}$;

\item
an annihilation amplitude ${\cal A}$.
\end{itemize}
The possible rescattering diagrams are not shown in Fig.\ref{fig:feynman}, one
can see Fig.12 of Ref.\cite{epjc6-451} for some relevant rescattering diagrams.

Following Refs.\cite{prd57-2752,gronau95}, the transition amplitudes for the
four $B \to \pi K$ decays can be written as
\beq
A(B^+ \to \pi^+ K^0)&=& P + c_d P^{C}_{EW} + {\cal A}\; , \label{eq:a1}\\
\sqrt{2} A(B^+ \to \pi^0 K^+)&=&
-\left [  P + T + C + P_{EW} + c_u P^{C}_{EW} + {\cal A} \right ]\; , \label{eq:a2}\\
A(B_d^0 \to \pi^- K^+)&=& -\left [ P + T + c_u P^{C}_{EW} \right ] \; ,\label{eq:a3} \\
\sqrt{2} A(B_d^0 \to \pi^0 K^0)& = &
 \left [ P - C \right ] - \left [ P_{EW} -c_d P^{C}_{EW} \right ]\;
 \label{eq:a4}
\eeq
where $c_u=2/3$ and $c_d=-1/3$ are the up- and down-type quark charges, respectively.
Because of the small ratio $|V_{us}V^*_{ub}|/|V_{ts}V^*_{tb}| \approx 0.02$,
the four $B \to \pi K$ decays are dominated by the QCD penguin $P$.
Because of the large top quark mass, we have also to care about the EW
penguins. The overall EW penguin amplitude should be ${\cal O}(10\%)$ that of the gluonic
penguin $P$, module group-theoretic factors\cite{gronau95}.
The EW penguins contribute in the color-suppressed form to $B^+ \to \pi^+ K^0$ and
$B_d^0 \to \pi^- K^+$ decays and are hence expected to play a minor role, whereas they
contribute to $B_d^0 \to \pi^0 K^0$ and $B^+ \to \pi^0 K^+$ decays in the color-allowed form
and may compete with tree-diagram-like topologies.
Approximately, the contribution from $P_{EW}^C$ should be smaller than its
color-allowed counterpart $P_{EW}$ by a factor of $0.2$ and is at
most a $5\%$ effect in $b \to s$ transition relative to the dominant QCD penguin contribution.

The relative sizes of the diagrams corresponding to the $\bar{b} \to \bar{u} u \bar{s}$
and $\bar{b} \to \bar{s}$ transitions at the quark level have been estimated\cite{gronau95}:
\beq
1 &:& |P|, \non
{\cal O}(\lambda) &:& |T|, |P_{EW}|, \non
{\cal O}(\lambda^2) &:& |C|, |P^C_{EW}|, \non
{\cal O}(\lambda^3) &:& |A|,
\label{eq:rsize}
\eeq
where the parameter $\lambda=0.22$ is used as a measure of the approximate relative sizes
of the various contributions. One can regard the above hierarchies as a simple estimation
\cite{gronau95} since a modest enhancement or suppression due to hadronic matrix element
for example can turn an effect of ${\cal O}(\lambda^n)$ into an effect of
${\cal O}(\lambda^{n\pm 1})$.

In Refs.\cite{epjc6-451,epjc11-93}, the  decay amplitudes of
$B\to \pi^+ K^0$ and $\pi^- K^+$ have been parametrized as follows
\footnote{ In the parameterization, the $SU(2)$ isospin symmetry of $u$ and $d$
quarks, the unitarity of the CKM matrix and the Wolfenstein parametrization of the CKM
matrix have been applied. }
\beq
A(B^+ \to \pi^+ K^0) &\equiv & {\cal P } = -\left ( 1-\frac{\lambda^2}{2} \right ) \lambda^2 A \left [
1 + \rho e^{i\theta} e^{i\gamma} \right ] {\cal P}_{tc}, \label{eq:ap0}\\
A(B_d^0 \to \pi^- K^+) &=& - \left [ {\cal P } + T + P_{EW}^C \right ],
\label{eq:amp}
\eeq
with
\beq
 {\cal P}_{tc} &\equiv & | {\cal P}_{tc}| e^{i\delta_{tc}}=\left (P_t -P_c \right )
 + \left (P_{EW}^{C(t)} -P_{EW}^{C(c)} \right ), \label{eq:ptc} \\
 \rho e^{i \theta}&=& \frac{\lambda^2 R_b}{1-\lambda^2/2}\left [1-\left(
   \frac{{\cal P}_{uc}+ {\cal A}}{{\cal P}_{tc}}\right )\right ],
   \label{eq:rhom}\\
T  &\equiv & |T| e^{i\delta_T} e^{i\gamma}, \label{eq:tm}\\
P_{EW}^C  &\equiv & - |P_{EW}^C| e^{i\delta_{ew}}, \label{eq:pewc}
\eeq
where $P_q$ and $P_{EW}^{C(q)}$ ($q \in \{ u,c,t\}$) denote contributions from
QCD penguin and color-suppressed electroweak penguin topologies with internal q
quarks, respectively.  ${\cal P}_{uc}$ is similar to ${\cal P}_{tc}$ in
Eq.(\ref{eq:ptc}), $\theta$, $\delta_{tc}$, $\delta_{T}$ and $\delta_{ew}$
are CP-conserving strong phases, and
\beq
A &=& 0.85 \pm 0.04, \ \ \lambda=0.221 \pm 0.002, \ \ R_b=0.38 \pm 0.08
\eeq
are the usual CKM factors\cite{buras01}.

The ratio $R$ and the corresponding "pseudo-asymmetry" $A_0$ then take the
form\cite{epjc6-451}
\beq
R&= &   1 + r^2 + \epsilon^2  -2r \epsilon \cos{(\delta-\Delta)}\cos{\gamma}
  - \frac{2r}{\omega}\left [  \cos{\delta}\cos{\gamma} + \rho
  \cos{(\delta-\theta)}\right  ] \non
   &&  +  \frac{2 \epsilon}{\omega} \left [\cos{\Delta} + \rho \cos{(\Delta-\theta)}
   \cos{\gamma} \right ], \label{eq:rm2} \\
A_0 &=& A_+ + \frac{2r}{\omega} \sin{\delta}\sin{\gamma}
+  \frac{2 \epsilon}{\omega}  \rho \sin{(\Delta-\theta)} \sin{\gamma}
+ 2r\epsilon \sin{(\delta-\Delta)}\sin{\gamma},  \label{eq:am2}
\eeq
where $A_+$ measures the direct CP violation in the decay $B^+ \to \pi^+ K^0$
\beq
A_+ &\equiv & \frac{BR({B}^+ \to \pi^+ K^0)- BR({B}^- \to \pi^-0 \overline{K^0})}
   {BR(B^+ \to \pi^+ K^0) + BR(B^- \to \pi^- \overline{K^0})}\non
  &=& - \frac{2\rho}{\omega^2}\sin\theta \sin\gamma \label{eq:ap}
\eeq
with
\beq
\omega= \sqrt{1 + 2\rho \cos{\theta}\cos{\gamma} + \rho^2}. \label{eq:omega}
\eeq
The parameters $r$ and $\epsilon $, as well as the CP-conserving strong phases
$\delta$ and $\Delta$, have been defined as follows
\beq
r &\equiv & \frac{|T|}{\sqrt{<|{\cal P}|^2>}}, \ \ \epsilon \equiv
\frac{|P_{EW}^C|}{\sqrt{<|{\cal P}|^2>}}, \label{eq:repsilon} \\
\delta &\equiv & \delta_T - \delta_{tc}, \ \
\Delta \equiv  \delta_{ew} - \delta_{tc}, \label{eq:delta1}
\eeq
with
\beq
<|{\cal P}|^2> \equiv \frac{1}{2} \left ( |{\cal P}|^2  + |\overline{{\cal P}}|^2 \right ).
\eeq
Here the $\overline{{\cal P}}$ is the CP-conjugate modes of ${\cal P}$ and obtained by performing the
substitution $\gamma \to - \gamma$.

For the decay $B \to \pi^0 K^0 $, one parametrization presented in Ref.\cite{epjc11-93} is
\beq
\sqrt{2} A(B_d^0 \to \pi^0 K^0) & \equiv & P_n
= -\left (  1-\frac{\lambda^2}{2}\right ) \lambda^2 A \left [ 1 +
\rho_n e^{i\theta_n} e^{i\gamma} \right ]{\cal P}_{tc}^n, \label{eq:a00}
\eeq
where ${\cal P}_{tc}^n$ and $\rho_n e^{i\theta_n}$ take the form
\beq
{\cal P}_{tc}^n &\equiv & |{\cal P}_{tc}^n| e^{i\delta_{tc}^n}, \non
\rho_n e^{i\theta_n} &=& \frac{\lambda^2 R_b}{1-\lambda^2} \left [
1- \left ( \frac{{\cal P}_{uc}- {\cal C} }{{\cal P}_{tc}^n}\right )\right ]
\eeq

By using the isospin symmetry of strong interactions for the $u$ and $d$ quarks
and the decay amplitude relations as given in Eqs.(\ref{eq:a1},\ref{eq:a4}), we here
parametrize the $B \to \pi^0 K^0$ decay in a new way
\beq
\sqrt{2} A(B_d^0 \to \pi^0 K^0) \equiv P_n =  {\cal P} - C - P_{EW}, \label{eq:a01}
\eeq
with
\beq
  C \equiv | C|e^{i \delta_{c}}e^{i \gamma} \ \ \ P_{EW}\equiv
   -|P_{EW}|e^{i \delta'_{ew}},
\eeq
where the term $C$ denotes the contributions due to the color-suppressed "tree" diagrams,
the quantity $P_{EW}$ includes contributions from the color-allowed EW penguin topologies,
and the $\delta_c$ and $\delta'_{ew}$ denote CP-conserving strong phases.

If  we  define the observables
\beq
r_1&\equiv & \frac{| C |}{\sqrt{<|{\cal P}|^2>}},  \ \ \ \
\epsilon_1  \equiv \frac{|P_{EW}|}{\sqrt{<|{\cal P}|^2>}} , \non
\delta_1 & \equiv & \delta_c-\delta_{tc} , \ \ \ \
\Delta_1 \equiv    \delta'_{ew}-\delta_{tc}\label{eq:r1d}
\eeq
we then find the expressions for $R_1$ and  $A_1$
\beq
R_1 &=& 1+ r_1^2+\epsilon_1^2+ \frac{2r_1}{\omega} \left [
\cos{\delta_1}\cos{\gamma}+\rho \cos{(\delta_1-\theta)} \right ] \non
   & & - \frac{2\epsilon_1}{\omega} \left [ \cos{\Delta_1}+\rho
   \cos{(\Delta_1-\theta)}\cos{\gamma} \right ] -2r_1 \epsilon_1
   \cos{(\delta_1-\Delta_1)}\cos{\gamma} \label{eq:r1}\\
A_1& = &  A_+ - \frac{2r_1}{\omega} \sin{\delta_1}\sin{\gamma}
  - \frac{2\epsilon_1}{\omega}  \rho \sin{(\Delta_1-\theta)} \sin{\gamma}
   +  2r_1 \epsilon_1 \sin{(\delta_1-\Delta_1)}\sin{\gamma},
\label{eq:a11}
\eeq
where the parameters $\rho$, $\theta$, $\omega$ and the CP asymmetry $A_+$ have been
defined previously. In Eqs.(\ref{eq:r1},\ref{eq:a11}), the electroweak penguin and
rescattering effects are taken into account in a general way.

From the estimated hierarchy between the different diagrams as given in Eq.(\ref{eq:rsize}),
we get to know that
\beq
r \approx \epsilon_1 \approx 0.2, \ \ r_1 \approx \epsilon \approx 0.04.\label{eq:rra}
\eeq
Evaluations based on the generalized factorization approach indicated
that\cite{prd57-2752,epjc11-93}
\beq
r|_{fact}=0.16 \pm 0.05, \ \ \epsilon|_{fact} =0.01-0.03.\label{eq:rrb}
\eeq

By direct calculations in the generalized factorization approach we find numerically that
\beq
r =0.14- 0.20, \ \ \epsilon =0.01-0.04,
\eeq
and
\beq
r_1= 0.001- 0.04, \ \ \epsilon_1=0.07 - 0.15 \label{eq:rrc}
\eeq
in the case of neglected rescattering effects.
Of course, above estimations may be affected severely by rescattering effects,
which is unfortunately still unknown at present.  A reliable
theoretical evaluation of $\rho$ is indeed very difficult and requires insights into
the dynamics of strong interactions. In Ref.\cite{epjc6-451}, Fleischer studied
the rescattering processes of the kind
\beq
B^+ &&\to \{ F_c^{(s)}\} \to \pi^+ K^0, \label{eq:rho1}\\
B^+ &&\to \{ F_u^{(s)}\} \to \pi^+ K^0, \label{eq:rho2}
\eeq
where $F_c^{(s)} \in \{ \overline{D^0} D_s^+, \overline{D^0} D_{s}^{*+}, \cdots\}$ and
$F_u^{(s)} \in \{ \pi^0 K^+, \pi^0 K^{*+}, \cdots\}$, and found that
(a) $\rho \approx 0$  if rescattering processes of type (\ref{eq:rho1})
played the dominant role in $B^+ \to \pi^+ K^0$ decay;
(b) $\rho ={\cal O}(10\%)$ if (\ref{eq:rho2}) is dominant; and
(c) $\rho ={\cal O}(\lambda^2 R_b) \approx 0.04$ if both  (\ref{eq:rho1}) and
(\ref{eq:rho2}) were similarly important.

\section{Bounds on $\gamma$ from $R$ and $A_0$: an update}    \label{sec-3}

In Refs.\cite{prd57-2752,epjc6-451,epjc11-93}, the strategies to extract the CKM angle $\gamma$
from the ratio $R$ have been studied.
Because of the refinement of the measured $R$ and the first measurement of $A_+$,
we here make an update for this approach.

Very recently, CLEO, BaBar and Belle collaboration reported their first measurement about
the CP violating asymmetries of $B \to  \pi^+ K^-, \pi^+ K^0$ and $\pi^0 K^+$ decays
\cite{babar2001,acp2001}, as listed in Table \ref{tab:acp}.
Although the measured CP-violating asymmetries of three $B \to \pi K$ decay modes have large
uncertainty and therefore are still consistent with zero,  we believe that they will be
measured with a good accuracy within one or two years. From the measured $R$ and
$\acp(B\pm \to \pi^\pm K^\mp)$, we find that
\beq
A_0 \equiv  \acp({B^\pm \to \pi^\pm K^\mp})\cdot R = -0.12 \pm 0.13.\label{eq:amdata}
\eeq
A recent theoretical calculation based on PQCD approach predicted that
$\acp(B_d^0 \to \pi^\pm K^\mp)\approx -0.19$ for $\gamma \sim 60^\circ$ \cite{sanda01},
which is consistent with the experimental measurements. It is reasonable for us to assume
that $|A_0| \lesssim 0.2$.

The ratio $R$ and the asymmetry $A_0$ as given in Eqs.(\ref{eq:rm2},\ref{eq:am2}) depend on
seven parameters: $r$, $\epsilon$, $\gamma$, $\rho$ and CP-conserving  strong phases
$\theta$, $\delta$ and $\Delta$.
Although parameters $r$ and $\epsilon$ can be fixed through theoretical
arguments, the parameter $\rho$ is most possibly smaller than $0.15$,
other four parameters still remain unknown.

\begin{figure}[t]%fig.2
\begin{minipage}[]{\textwidth}
\centerline{\epsfxsize=0.8\textwidth \epsffile{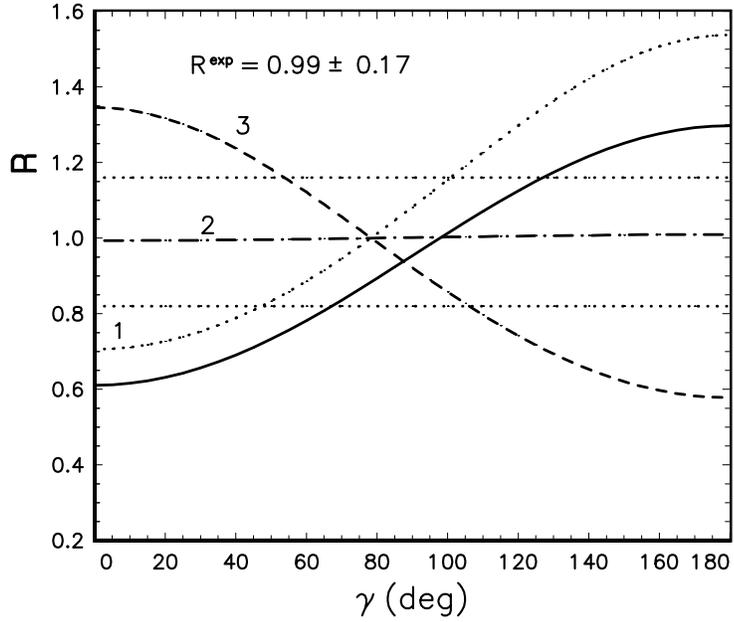}}
\vspace{-20pt}
\caption{The dependence of $R$ on $\gamma$ for $\rho=0.1$,
$\epsilon=0.04$, $r=0.2$, while assuming
$( \theta, \delta, \Delta)= 0^\circ$ (curve 1 ), $90^\circ$ (curve 2) and
$180^\circ $ (curve 3 ). The solid curve is the standard model
prediction based on the generalized factorization approach. The band
between two dots lines corresponds to the data: $R=0.99\pm 0.17$. }
\label{fig:rmsm1}
\end{minipage}
\end{figure}

In Fig.\ref{fig:rmsm1} we show the dependence of $R$ on the angle $\gamma$ for $\rho=0.1$,
$\epsilon=0.04$, $r=0.2$, while assuming
$( \theta, \delta, \Delta)= 0^\circ$ (curve 1), $90^\circ$ (curve 2) or $180^\circ $ (curve 3).
The solid curve corresponds to the standard model prediction obtained by
employing the generalized factorization approach and using the input parameters as
specified in Ref.\cite{x10326}. The band between two horizontal dots lines
shows the experimental measurement: $R^{exp} =0.99 \pm 0.17$.
From this figure, one can see that
\begin{itemize}
\item
The ratio $R$ in the generalized factorization approach has a
similar dependence on $\gamma$ with the ratio $R$ as given in Eq.(\ref{eq:rm2})
in the case of  $( \theta, \delta, \Delta)= 0^\circ$.

\item
The ranges of $\gamma < 45^\circ$ and $\gamma > 120^\circ$ can be excluded for extreme
values ($0^\circ$ or $180^\circ$) of three strong phases. But such constraint on $\gamma$
will disappear for $\theta = \delta = \Delta = 90^\circ$. In other words, no
bounds on $\gamma$ can be extracted directly from the $R-\gamma$ plane due
to our ignorance of the strong phases.
\end{itemize}

As shown in Ref.\cite{epjc6-451}, however,  the observable $A_0$ allow us to eliminate
the strong phase $\delta$ in the expression of $R$. By assuming both the parameter
$r$ and the strong phase $\delta$ in the expression of $R$ as "free" parameters,
one found the minimal value of $R$ as follows \cite{epjc6-451}
\beq
R_{min} = \kappa \sin^2\gamma + \frac{1}{\kappa} \left (   \frac{A_0}{2 \sin\gamma}
\right )^2,  \label{eq:rmmin}
\eeq
with
\beq
\kappa = \frac{1}{\omega} \left [  1 + 2(\epsilon \omega)\cos\Delta
+ (\epsilon \omega)^2\right ]. \label{eq:kappa}
\eeq
where $\omega$ has been given in Eq.(\ref{eq:omega}). Now $R_{min}$ is independent of
both $r$ and $\delta$, the effects of
electroweak penguin and rescattering are included through the parameters
$\epsilon$,  $\rho$ and $\theta$ appeared in $\kappa$ and $\omega$. By comparing the plots
of $R_{min}-\gamma$ with the measured $R$ as
given in Eq.(\ref{eq:data}), one may draw the bounds on the CKM angle $\gamma$.

\begin{figure}[t]%fig.3
\begin{minipage}[]{\textwidth}
\centerline{\epsfxsize=0.8\textwidth \epsffile{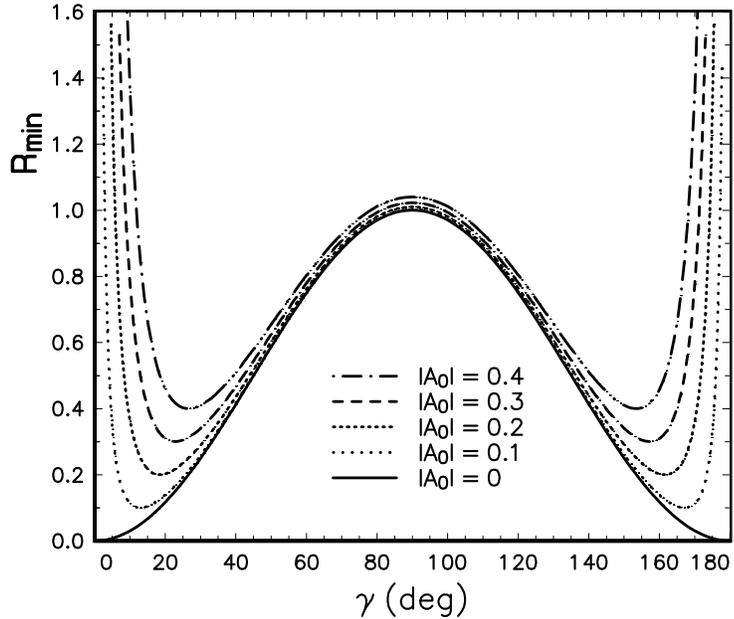}}
\vspace{-20pt}
\caption{The dependence of $R_{min}$ on $\gamma$ for $|A_0|=0, 0.1, 0.2, 0.3$
and $0.4$ in case of neglected EW penguin and rescattering effects.}
\label{fig:rma}
\end{minipage}
\end{figure}

As a first approximation, we neglect the rescattering and electroweak penguin effects
(i.e. setting $\rho=\epsilon =0$ ). The value of $R_{min}$ therefore depends on $A_0$
and $\gamma$ only.
The $\gamma-$dependence of $R_{min}$ for $A_0=0, 0.1, 0.2, 0.3$ and $0.4$
is shown in Fig.\ref{fig:rma}, which is identical with
the Fig.1 of Ref.\cite{epjc6-451}. From Fig.\ref{fig:rma} we get to know
that if $R$ is found to be smaller than $1$, the values of $\gamma$ implying $R_{min} >
R $ would be excluded. The current measured value of
$R$ is unfortunately very close to unit, it is also unlikely to become smaller
than $1$ when more B decay events are available. The bound on $\gamma$ from measurement of
$R$ is therefore not as promising as three years ago.

\begin{figure}[t]%fig.4
%\vspace{-40pt}
\begin{minipage}[]{\textwidth}
\centerline{\epsfxsize=0.8\textwidth \epsffile{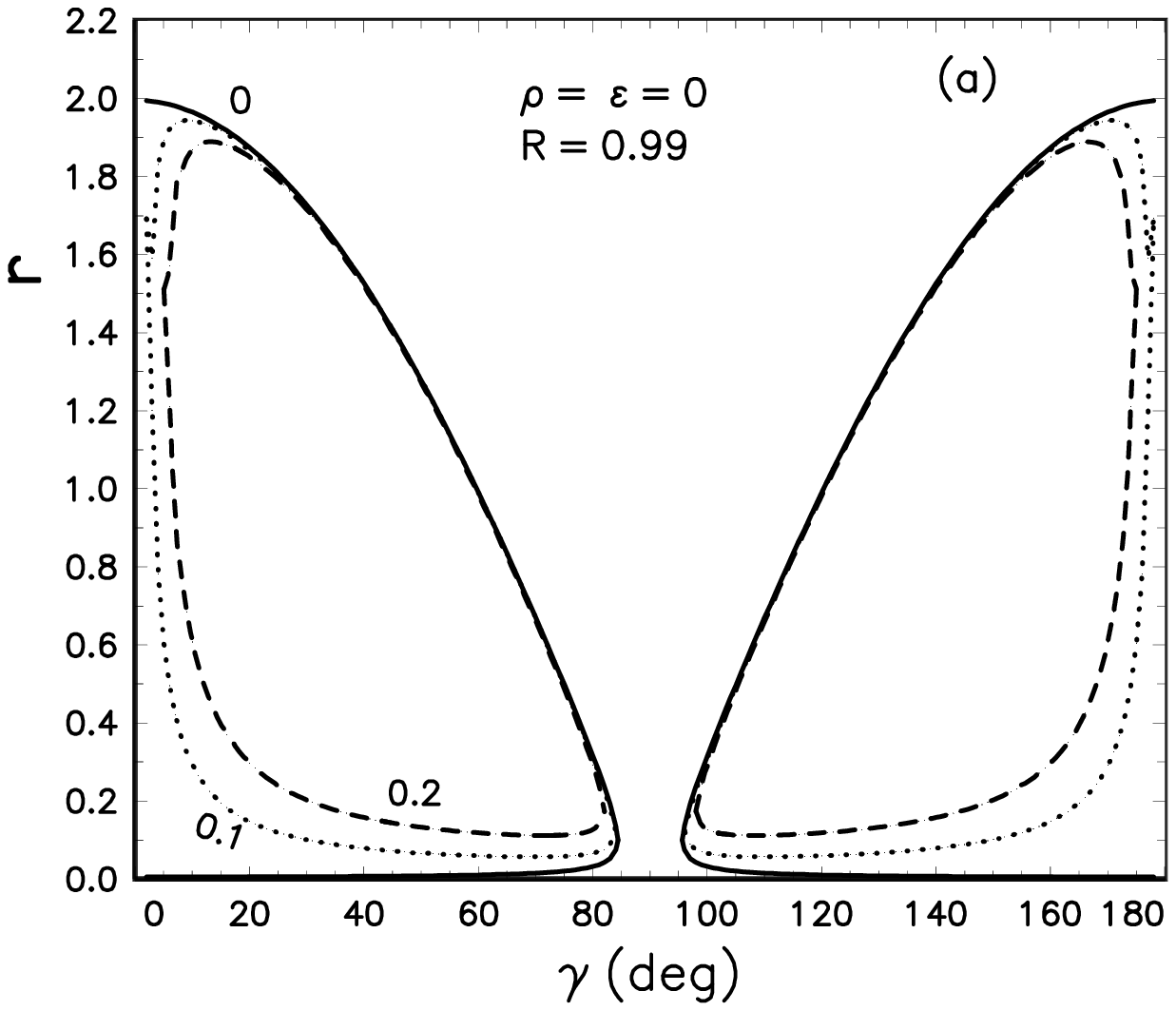}}
\vspace{-20pt}
\centerline{\epsfxsize=0.8\textwidth \epsffile{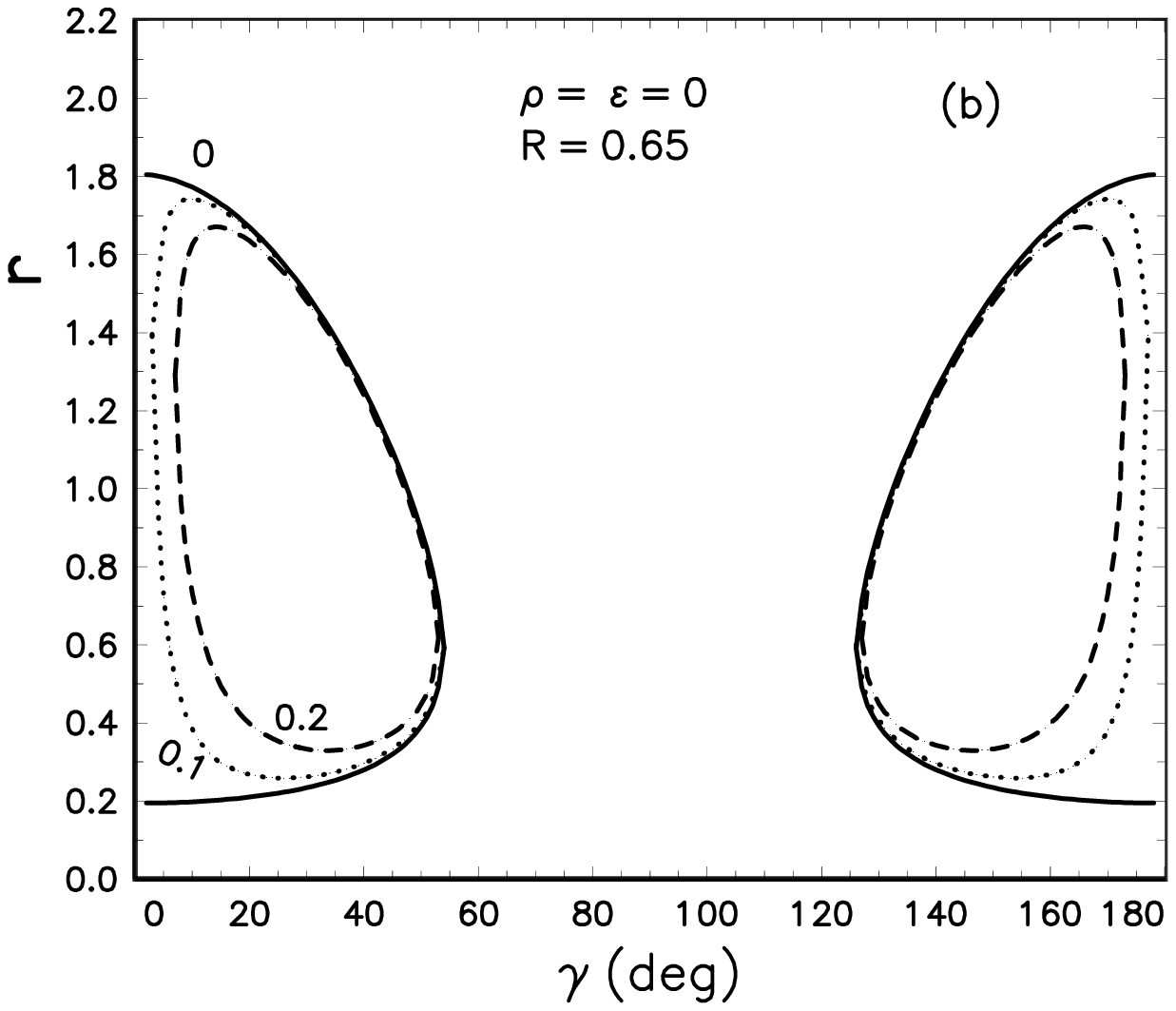}}
\vspace{-20pt}
\caption{The dependence of $r$ on the angle $\gamma$ for $R=0.99$ (a) and $ 0.65$ (b),
and for $|A_0| = 0$ (solid curve), $0.1$ (dots curve) and  $0.2$ (short-dashed curve)
in the case of $\rho=\epsilon=0$.}
\label{fig:rmc}
\end{minipage}
\end{figure}

For given $A_0$, the dependence of $r$ on the angle $\gamma$ is of the form \cite{epjc6-451}
\beq
r= \sqrt{ \left (R + \cos^2\gamma \right )
\pm 2 \sqrt{\cos^2\gamma \left (R -  \frac{A_0^2}{4\sin^2\gamma} \right )} }\label{eq:ra0}
\eeq
in the case of  $\rho=\epsilon=0$. Fig.\ref{fig:rmc}a show such dependence for $R=0.99$,
and $A_0=0$ (solid curve), $0.1$ (dots curve), $0.2$ (short-dashed curve).
Fig.\ref{fig:rmc}b shows the same dependence of $r$ on the angle
$\gamma$ but for $R=0.65$ as being used in Ref.\cite{epjc6-451}. The contours as shown
in Figs.(\ref{fig:rmc}a,\ref{fig:rmc}b) are rather different. For $R=0.65$, the
value of $r$ can not be smaller than $0.2$. For $R=0.99$, $r \approx 0.15$ as
indicated by theoretical calculations based on "factorization" is natural.
For given $R$ and $A_0$, the allowed ranges of $\gamma$ could be read off from the figures
if the parameter $r$ can be fixed by using an additional input.
For $|A_0|=0.2$, $R=0.99$ and $r=0.2$, for example, the  regions
\beq
0^\circ  \leq  \gamma \leq 32^\circ, \ \ 82^\circ \leq \gamma \leq 98^\circ,
\ \  and \ \ 150^\circ \leq \gamma \leq 180^\circ
\label{eq:gamma1}
\eeq
should be excluded. The above constraints are consistent
with the limit on $\gamma$ obtained from the global fit \cite{he2001}:
$43^\circ < \gamma < 87^\circ$ at the 95\% C.L.

For more details of  $\rho$ and $\epsilon$ dependence of $R_{min}$, one can see the
original papers\cite{epjc6-451,epjc11-93}. The  new measurement of $R$ and $A_0$
do not affect previous discussions.

\section{Bounds on $\gamma$ from $R_1$ and $A_1$}    \label{sec-4}

Analogous to the cases of $R$ and $A_0$\cite{prd57-2752,epjc6-451,epjc11-93},
the observables $R_1$ and $A_1$ may also lead to interesting bounds on $\gamma$.
By comparing  the expressions of $R_1$ and $A_1$ with those of $R$ and $A_0$,
we find three special features
\begin{itemize}
\item
Between the observables $(R_1, A_1)$ and $(R, A_0)$, there is a direct transformation
relation:  $r \to -r_1$ and $\epsilon \to -\epsilon_1$.

\item
The parameter $r_1$ which describes the contributions of "color-suppressed"
tree diagram is small in size: its "factorized" value is $(r_1)_{fact}=0.001-0.04$
and can be neglected.

\item
The parameter $\epsilon_1$ which describes the contributions of "color-allowed"
electroweak penguins, however, may be large in size as given in Eq.(\ref{eq:rrc}),
and usually can not be neglected.

\end{itemize}

Like the ratio $R$ and the asymmetry $A_0$, $R_1$ and  $A_1$ also depend on
seven parameters: $r_1$, $\epsilon_1$, $\rho$,  CKM angle $\gamma$ and the strong phases
$\theta$, $\delta_1$ and $\Delta_1$ as defined in Eqs.(\ref{eq:rhom},\ref{eq:r1d}).
The parameters $r_1$ and $\epsilon_1$ can be fixed through theoretical
arguments. If one can neglect or treat $\rho$ parameter and three strong phases
properly, one may determine or put constraint on the angle $\gamma$ from the measured
$R_1$.

\begin{figure}[t]%fig.5
\begin{minipage}[]{\textwidth}
\centerline{\epsfxsize=0.8\textwidth \epsffile{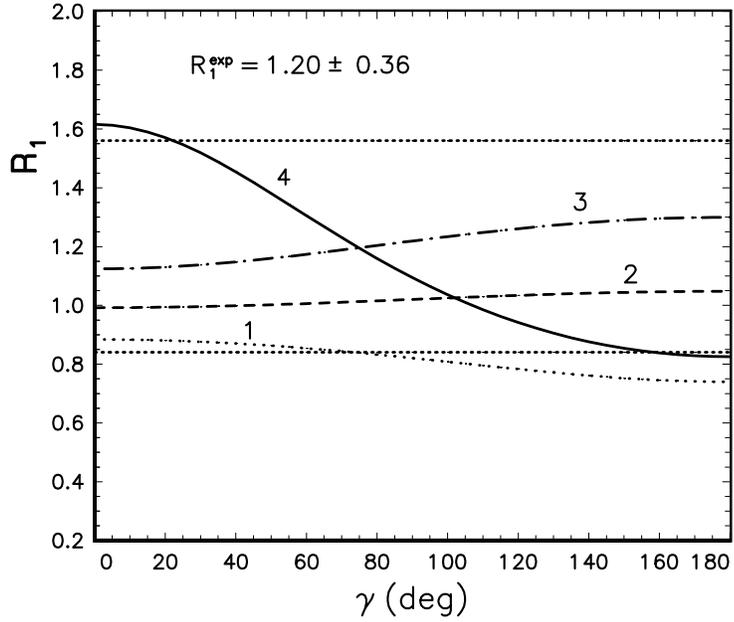}}
\vspace{-20pt}
\caption{The dependence of $R_1$ on $\gamma$ for $\rho=0.1$,
$r_1=0.04$, $\epsilon_1=0.1$, while assuming
$( \theta, \delta, \Delta)= 0^\circ$ (curve 1 ), $90^\circ$ (curve 2) and
$180^\circ $ (curve 3 ). The solid curve is the standard model
prediction of $R_1$ by employing the generalized factorization approach. The band
between two dots lines corresponds to the data: $R_1=1.20\pm 0.36$. }
\label{fig:r1sm1}
\end{minipage}
\end{figure}

In Fig.\ref{fig:r1sm1} we show the general dependence of the ratio $R_1$ in Eq.(\ref{eq:r1})
on the angle $\gamma$ for $\rho=0.1$, $r_1=0.04$, $\epsilon_1=0.1$, while assuming
$( \theta, \delta_1, \Delta_1)= 0^\circ$ (curve 1), $90^\circ$ (curve 2) or $180^\circ $ (curve 3).
The solid curve corresponds to the standard model prediction of $R_1$  obtained by
employing the generalized factorization approach and using the input parameters as
specified in Ref.\cite{x10326}. The band between two horizontal dots lines
shows the data: $R_1^{exp} =1.20 \pm 0.36$.
Obviously no constraint on the angle $\gamma$ can be obtained by comparing the measured
$R_1$ with the theory directly. It seems that the current data prefer large
strong phases (curve 3).

\begin{figure}[t]%fig.6
\begin{minipage}[]{\textwidth}
\centerline{\epsfxsize=0.8\textwidth \epsffile{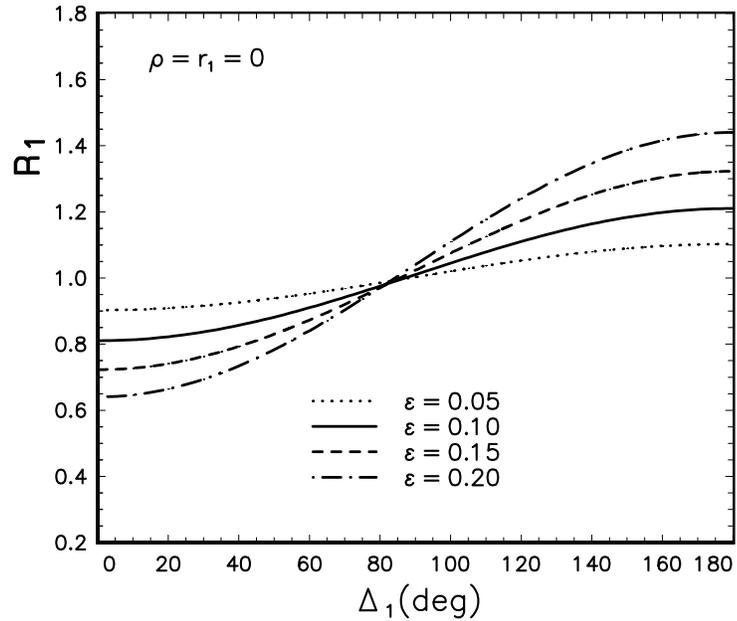}}
\vspace{-20pt}
\caption{The dependence of $R_1$ on the strong phase $\Delta_1$
for $\rho=r_1=0$, and $\epsilon_1=0.05, 0.10, 0.15$ and $0.2$. }  \label{fig:r1da}
\end{minipage}
\end{figure}

In case of neglected rescattering and the color-suppressed tree diagrams, the expression
of $R_1$ in Eq.(\ref{eq:r1}) can be greatly reduced into the form
\beq
R_1 = 1 + \epsilon_1^2 -2 \epsilon_1 \cos\Delta_1. \label{eq:r1m}
\eeq
Now it depends on two parameters only. If one can fix the value of $\epsilon_1$
from theoretical arguments, the measured $R_1$ will imply  limits on the strong
phase $\Delta_1$. In Fig.\ref{fig:r1da} we show the dependence of $R_1$ on the strong phase
$\Delta_1$ for various values of $\epsilon_1$. For given $\epsilon_1=0.2$, the lower limit on
$\Delta_1$ can be read off directly from this figure
\beq
\Delta_1  \geq  115^\circ, \label{eq:limitd1}
\eeq
for $R_1=1.2$. In other words, the measured $R_1$ prefers $\Delta_1 \geq 90^\circ$,
i.e. $\cos\Delta_1 < 0$.

Following the same procedure of Ref.\cite{epjc6-451}, one can eliminate the strong phase
$\delta_1$ in Eq.(\ref{eq:r1}) and find the minimal value of the ratio $R_1$.
For this purpose, we rewrite Eq.(\ref{eq:r1}) and Eq.(\ref{eq:a11}) as
\beq
R_1&=& R_0 + 2r_1 \left (\sl h \cos{\delta_1}+ \sl k \sin{\delta_1} \right
)+r_1^2,    \label{eq:e24}\\
A&=&\left ( B\sin{\delta}- C\cos{\delta} \right )r_1 ,     \label{eq:e25}
\eeq
where the quantities
\beq
R_0&=& 1 - 2\frac{\epsilon_1}{\omega}\left [\cos{\Delta}+\rho \cos{(\Delta_1-\theta)}
    \cos{\gamma} \right ]+\epsilon_1^2,  \label{eq:r0}\\
\sl h &=&\frac{1}{\omega}\left (\cos{\gamma}+\rho
    \cos{\theta}\right )-\epsilon_1 \cos{\Delta_1}\cos{\gamma},\\
\sl k &=& \frac{\rho}{\omega}\sin{\theta}-\epsilon_1
\sin{\Delta_1}\cos{\gamma}, \\
A&=&\frac{A_1-A_+}{2\sin{\gamma}}+\frac{\epsilon_1
    \rho}{\omega} \sin{(\Delta_1-\theta)}, \\
B&=&-\left (\frac{1}{\omega}-\epsilon_1 \cos{\Delta_1} \right ), \\
C&=& \epsilon_1 \sin{\Delta_1}, \label{}
\eeq
are independent of $r_1$. From Eq.(\ref{eq:e25}), we get
\beq
\sin{\delta_1}&=&\frac{AB\pm C
    \sqrt{(B^2+C^2) r_1^2-A^2}}{(B^2+C^2) r_1 }, \non
\cos{\delta_1}&=& \frac{-AC\pm B \sqrt{(B^2+C^2) r_1^2-A^2}}{(B^2+C^2 ) r_1},
\eeq
and then eliminate the strong phase $\delta_1$ in Eq.(\ref{eq:e24}):
\beq
R_1 =R_0 -AD\pm E\sqrt{(B^2+C^2) r_1^2-A^2}+r_1^2,     \label{eq:e31}
\eeq
with
\beq
    D=2\left ( \frac{ h C- k B}{B^2+C^2}\right ), \ \ \
    E=2\left ( \frac{ h B+ k C}{B^2+C^2}\right )
\eeq
Treating now $ r_1$ in Eq.(\ref{eq:e31}) as a free variable, we find the minimal value
of $R_1$
\beq
(R_1)_{min} = t \sin^2{\gamma}+\frac{1}{ t}\left( \frac{A_1}{2\sin{\gamma}}\right)^2
    \label{eq:r1min}
\eeq
with
\beq
t=\frac{1}{\omega^2}\left [1 - 2\epsilon_1 \omega \cos{\Delta_1}
                             + \epsilon_1^2\omega^2 \right ]. \label{eq:tt}
\eeq
This is an exact formulae derived without any approximation. The effects of electroweak
penguin and rescattering processes are included through the parameter $\epsilon_1$,
$\rho$ and $\theta$, respectively.

\begin{figure}[t]%fig.7
\begin{minipage}[]{\textwidth}
\centerline{\epsfxsize=0.8\textwidth \epsffile{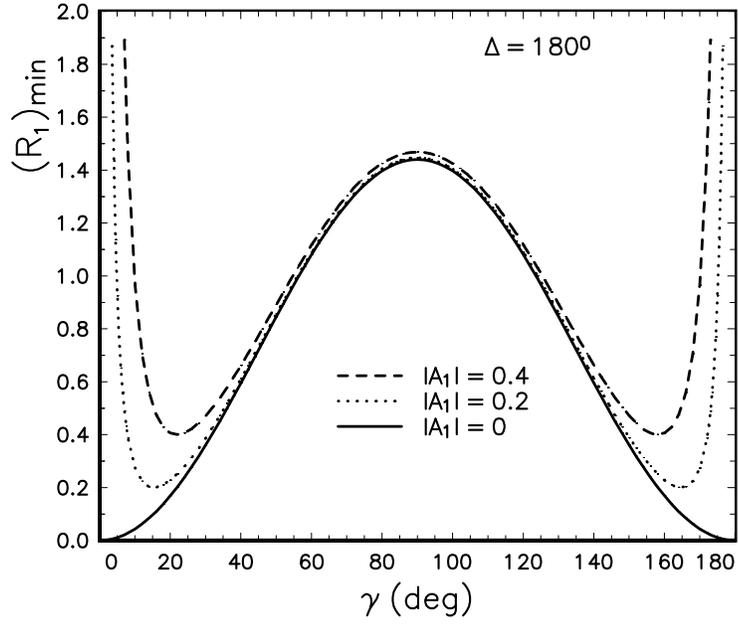}}
\vspace{-20pt}
\caption{The dependence of the minimal value of $R_1$ on the angle $\gamma$
for $\rho=0$, $\epsilon_1=0.2$, $\Delta_1=180^\circ$ and $|A_1|=0,0.2$ and $0.4$. }
\label{fig:r1min1}
\end{minipage}
\end{figure}

\begin{figure}[t]%fig.8
\begin{minipage}[]{\textwidth}
\centerline{\epsfxsize=0.8\textwidth \epsffile{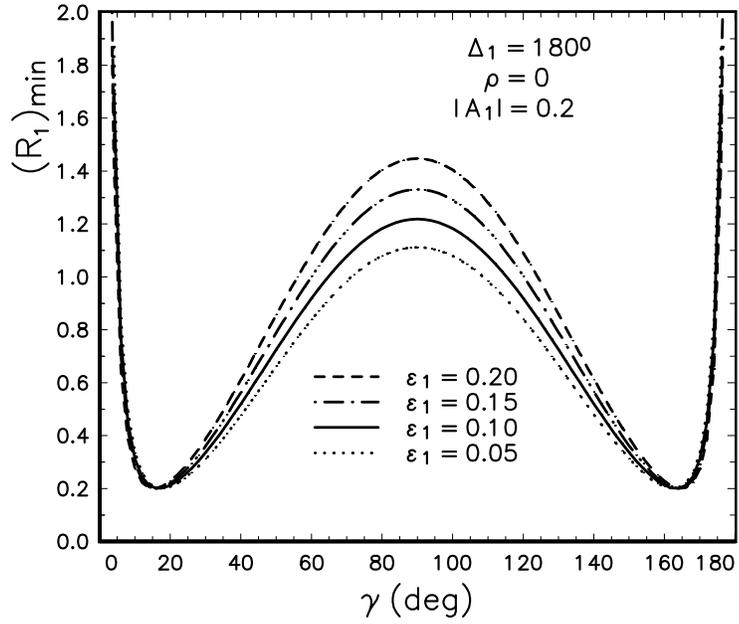}}
\vspace{-20pt}
\caption{The dependence of the minimal value of $R_1$ on the angle $\gamma$
for $\rho=0$, $A_1=0.2$,  $\Delta_1=180^\circ$, $\epsilon_1=0.05$, $0.10$, $0.15$
and $0.2$. }
\label{fig:r1min3}
\end{minipage}
\end{figure}

\begin{figure}[t]%fig.9
%\vspace{-40pt}
\begin{minipage}[]{\textwidth}
\centerline{\epsfxsize=0.8\textwidth \epsffile{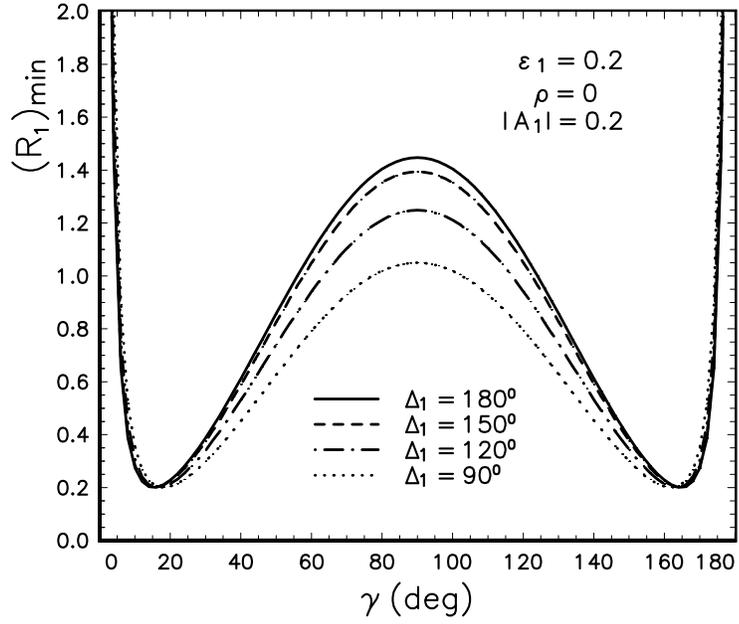}}
\vspace{-20pt}
\caption{The dependence of the minimal value of $R_1$ on the angle $\gamma$
for $\rho=0$, $\epsilon_1=0.2$, $|A_1|=0.1$, $\Delta_1=0^\circ$, $45^\circ$,
$90^\circ$, $135^\circ$  and $180^\circ$. }
\label{fig:r1min2}
\end{minipage}
\end{figure}

For $B \to \pi^+ K^0$ and $\pi^+ K^-$ decays the EW penguin contributes in  the
"color-suppressed" form only and therefore play a minor role.
For $B \to \pi^0 K^0$ decay, however, the "color allowed" electroweak penguin is important
and should be taken into account. This is the main difference between two sets of observables
$(R, A_0)$ and $(R_1, A_1)$. For $\rho=0$ and $\epsilon_1 \neq 0$, the minimal value of
$R_1$ will depend on the angle $\gamma$, the parameter $\epsilon_1$, the pseudo-asymmetry $A_1$
and the strong phase $\Delta_1$, as illustrated in
Figs.(\ref{fig:r1min1},\ref{fig:r1min3},\ref{fig:r1min2}).

Fig.\ref{fig:r1min1} shows the dependence of $\rmin$ on the angle $\gamma$ for
$\rho=0$, $\epsilon_1=0.2$, $\Delta_1=180^\circ$ (maximal effect) and $|A_1|=0$,
$0.2$ and $0.4$. From Fig.\ref{fig:r1min1}, we find that
\begin{itemize}
\item
Due to the contribution from the  "color allowed" electroweak penguin, the value of
$\rmin$ can be larger than unit now:
\beq
R_1 \geq t \sin^2\gamma
\eeq
for $A_1=0$. Here the function $t$ is larger than unit if $\cos\Delta_1 < 0$.
The values of $\gamma$ implying $(R_1)_{min} > R_1$ would be excluded.
Numerically, the ranges around $\gamma=90^\circ$, i.e.
\beq
&& 58^\circ \leq \gamma \leq 122^\circ, \non
&& 65^\circ \leq \gamma \leq 115^\circ, \non
&& 80^\circ \leq \gamma \leq 100^\circ,
\eeq
would be excluded for $R_1 =1, 1.2 $ and $1.4$, respectively.
According to current experimental measurements, $R_1\approx 1.2$ is indeed natural, the
corresponding bounds on the angle $\gamma$ are thus practical and interesting.
Such bounds are also complimentary to the limits from global fit.

\item
Around $\gamma=90^\circ$, the bound on $\gamma$ is approximately
independent of $A_1$.  The excluded regions around
$\gamma=0^\circ$ and $180^\circ$, however,  depend on the values of $A_1$.

\end{itemize}

For the minimal value of $R_1$, the EW penguin contribution is included through
$t = 1- 2 \epsilon_1 \cos\Delta_1 + \epsilon_1^2$ in the case of $\rho=0$.
Fig.\ref{fig:r1min3} shows the dependence of $\rmin$ on the angle $\gamma$ for
$\rho=0$, $|A_1|=0.2$, $\Delta_1=180^\circ$,  $\epsilon_1=0.05, 0.10, 0.15$ and $0.20$.
It is easy to see that $\rmin$ has a moderate dependence on the value of $\epsilon_1$.
Using  $\epsilon_1= 0.1$ and $R_1=1.20$, the range of $83^\circ \leq \gamma \leq 97^\circ$
could be excluded.

Fig.\ref{fig:r1min2} shows the dependence of $\rmin$ on the angle $\gamma$ for
$\rho=0$, $|A_1|=0.2$, $\epsilon_1=0.2$,  $\Delta_1= 90^\circ, 120^\circ$, $150^\circ$
and $180^\circ$. Obviously, $\rmin$ and thus the bound on $\gamma$ has a strong dependence
on phase $\Delta_1$. The bound given in Eq.(\ref{eq:limitd1}) is the first
limit on $\Delta_1$ from the measured $R_1$, but its uncertainty is also large.
To get a reliable bound on $\gamma$ from this strategy, one has to determine the
value of $\Delta_1$ with a reasonable accuracy.

\begin{figure}[t]%fig.10
%\vspace{-40pt}
\begin{minipage}[]{\textwidth}
\centerline{\epsfxsize=0.8\textwidth \epsffile{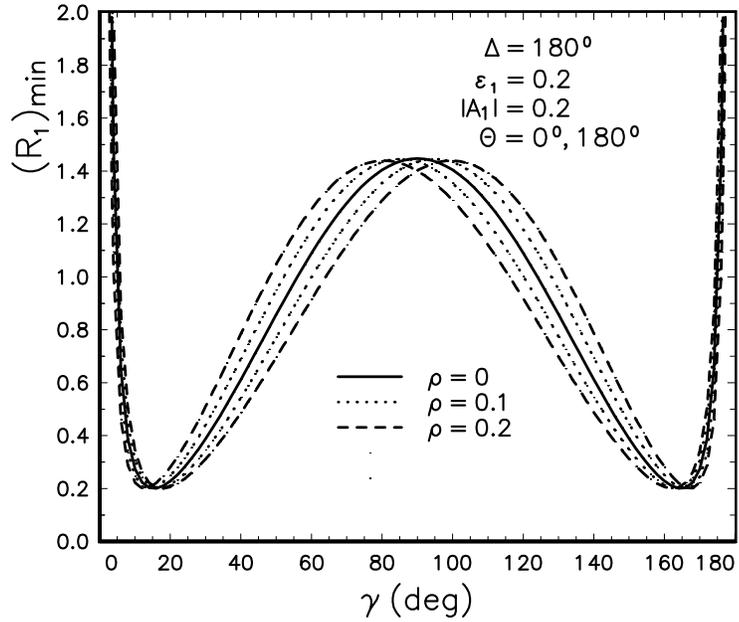}}
\vspace{-20pt}
\caption{The rescattering effects on $\rmin$ for $\rho=0, 0.10$ and $0.15$, while assuming
$\epsilon_1=0.1$, $\Delta_1=180^\circ$, $|A_1|=0.2$, $\theta\in \{0^\circ, 180^\circ \}$.
The curves for a given value of $\rho$ correspond to $\theta\in \{0^\circ, 180^\circ \}$
and show the maximum shift from $\rho=0$.}
\label{fig:r1min4}
\end{minipage}
\end{figure}

Now we check the effects of the rescattering. For the minimal value of $R_1$, the
rescattering effects are included  through $\omega = \sqrt{1 + 2\rho \cos\theta \cos\gamma
+\rho^2}$, and maximum for $\theta=0^\circ$ or $180^\circ$.
In Fig.\ref{fig:r1min4} we show the dependence of $\rmin$ on the angle $\gamma$
for $\rho=0$, $0.1$, and $0.2$, while assuming $\epsilon_1=0.2$, $\Delta_1=180^\circ$,
$|A_1|=0.2$, and  $\theta=0^\circ$ or $180^\circ$.
As shown in Fig.\ref{fig:r1min4}, $\rmin$ has a weak dependence on
$\rho$ only:  its maximum around $\gamma=90^\circ$ is almost independent of $\rho$.
The uncertainty of the bound on the  angle $\gamma$ for a given $R_1$ is at most
$10^\circ$ in the range of $\rho=0 - 0.2$.

\begin{figure}[t]%fig.6
%\vspace{-40pt}
\begin{minipage}[]{\textwidth}
\centerline{\epsfxsize=0.8\textwidth \epsffile{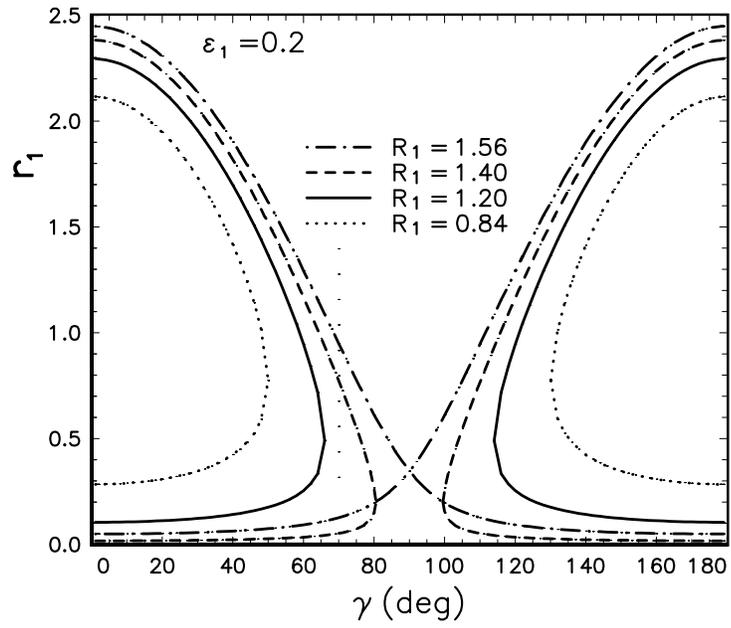}}
\vspace{-20pt}
\caption{The allowed regions of the parameter $r_1$ for $\rho=0$, $\epsilon_1=0.2$,
$\Delta=180^\circ$, and $R_1^{exp}=0.84, 1.20, 1.40$ and $1.56$.}
\label{fig:r1b4}
\end{minipage}
\end{figure}

According to the estimated hierarchy and direct calculation in generalized
factorization approach, the parameter $r_1$ should be very small: $0 \leq r_1 \leq
0.04$. By treating the $\delta_1$ in Eq.(\ref{eq:e24}) as a free parameter, on
the other hand, one can also put the lower and upper bounds on $r_1$ from the
measured $R_1$
\beq
(r_1)_{min}^{max} = \left |
\sqrt{R_0 - t \sin^2\gamma} \pm \sqrt{R_1 - t \sin^2\gamma}
 \right |. \label{eq:r1limit}
\eeq
where $R_0$ and $t$ have been given in Eqs.(\ref{eq:r0},\ref{eq:tt}).
Fig.\ref{fig:r1b4} shows the allowed regions of $r_1$ for $\rho=0$,
$\epsilon_1 =0.2$, $\Delta_1 =180^\circ$ and for various values of $R_1$
corresponding to its currently allowed experimental range $R_1=1.20 \pm 0.36$.
From this figure, one can see that
\begin{itemize}
\item
Small values of $R_1$ requires large values of $r_1$. For $R_1 =0.84$, for example,  the minimal
value of $r_1$ is $0.28$, which is much larger than the  theoretical estimations: $r_1 \sim 0.04$.
For $R_1 \geq 1.2$, however, the minimal values of $r_1$ become compatible with
the theoretical estimations.

\item
If $r_1$ could be fixed by using an additional input, the bounds on $\gamma$ can be read
off directly from Fig.\ref{fig:r1b4}. For $R_1=1.40$ and $r_1=0.05$, for
example, the range of
\beq
78^\circ \leq \gamma \leq 102^\circ
\eeq
could be excluded.

\end{itemize}

\section{Conclusions}

In this paper we defined and studied a new "mixed" system,  $B^+ \to \pi^+ K^0$ and
$B_d^0 \to \pi^0 K^0$ decays, to extract or constrain the CKM angle
$\gamma$ from the measured ratio $R_1$ and the corresponding "pseudo-asymmetry"
$A_1$. We also made an update for the constraints on the angle $\gamma$ from the
observables $R$ and $A_0$.

In the parameterization, the $SU(2)$ isospin symmetry of strong interactions for $u$ and $d$
quarks, the unitarity of the CKM matrix and the Wolfenstein parametrization of the CKM
matrix have been applied. From the theoretical calculations and currently available experimental
measurements, we found the following results:
\begin{itemize}
\item
The measured value of $R$ is now very close to unit, the bound on the angle $\gamma$
from measurement of $R$ is therefore not as promising as three years ago.
If $A_0$ is measured with good accuracy and $r$ can be fixed by theoretical arguments,
however, some bounds on $\gamma$ are still possible.
For $\rho=\epsilon=0$, $R=0.99$, $|A_0|=0.2$ and $r=0.2$, for example, the allowed regions
of angle $\gamma$ can be read off from Fig.\ref{fig:rmc}
\beq
30^\circ  \leq  \gamma \leq 82^\circ, \ \ and \ \ 98^\circ \leq \gamma \leq 150^\circ.
\eeq

\item
For $B_d^0 \to \pi^0 K^0$ decay, the color-allowed EW penguin play an important role.
By direct calculations in the generalized factorization approach we find that
\beq
r_1= 0.001- 0.04, \ \ \epsilon_1=0.07 - 0.15
\eeq
in the case of neglected rescattering effects, which agrees well with
the estimated hierarchy of different Feynman diagrams.

\item
As shown in Fig.\ref{fig:r1da},  the measured $R_1$ implies a limit on the strong phase
$\Delta_1$ in case of neglected rescattering and the color-suppressed tree
diagrams. For given $\epsilon_1=0.2$ and $R_1=1.2$, the lower limit on
$\Delta_1$ is $\Delta_1  \geq  115^\circ$. This  is the first
limit on $\Delta_1$ from the measured $R_1$, but its uncertainty is still large.

\item
Due to the contribution from the  "color allowed" electroweak penguin, the minimal value
of $R_1$ can be larger than unit, as illustrated in Fig.\ref{fig:r1min1}.
Using  $\epsilon_1=0.2$ and $R_1=1.20$, the
range of $65^\circ \leq \gamma \leq 115^\circ$ could be excluded, which is also
approximately independent of $A_1$. These bounds on the angle $\gamma$ are interesting and
complimentary to the limits from global fit.

\item
The bound on $\gamma$ has a moderate dependence on the value of $\epsilon_1$.
Using  $R_1=1.20$ and $\epsilon_1= 0.1$,  the range of $83^\circ \leq \gamma \leq
97^\circ$ could be excluded.

\item
The bound on $\gamma$ has a strong dependence on the strong phase $\Delta_1$.
For $\Delta_1 < 90^\circ$, for example, the bound may disappear.
To get a reliable bound on $\gamma$ from this strategy, one has to determine the
value of $\Delta_1$ with a reasonable accuracy.

\item
The bound on $\gamma$ has a weak dependence on the rescattering effects only.
The uncertainty of the bound on the  angle $\gamma$ for a given $R_1$ is at most
$10^\circ$ in the range of $\rho=0 - 0.2$.

\item
The bounds on $\gamma$ can be read off directly from $r_1 - \gamma$ plane if
$r_1$ could be fixed by using an additional input.

\end{itemize}

\section*{ACKNOWLEDGMENTS}
Z.J.~Xiao acknowledges the support by the National Natural Science Foundation of
China under Grant No.10075013, and by the Research Foundation of Nanjing Normal
University under Grant No.2001WLXXGQA916.

%%%%%%%%%%%%%%%%%%%%%%%%%%%%%%%%%%%%%%%%%%%%%%%%%%%%%%%%%%%%%%%%%%%%%%%%%%%%%%

\newpage
\begin{table}[t]
\begin{center}
\caption{Measurements of CP-violating asymmetry as reported by CLEO, BaBar and Belle
Collaborations. The numbers in second entries are the $\acp$ at $90\%$ C.L.
The last column lists the average.}
\label{tab:acp}
\vspace{0.2cm}
\begin{tabular} {|l|l|l|l|l|}  \hline
Channel & CLEO & BaBar & Belle & Average  \\ \hline
$\acp(B^\pm \to \pi^0 K^\pm)$&$-0.29 \pm 0.23$ & $0.00 \pm 0.18 \pm 0.04$ &$-0.06^{+0.22}_{-0.20} $ & $-0.10 \pm 0.12$ \\
 & $[-0.67, 0.09]$ &$[-0.30, 0.30]$ & $[-0.40, 0.36]$& \\
$\acp(B\to \pi^\mp K^\pm)$ &$-0.04 \pm 0.16$&$-0.19 \pm 0.10 \pm 0.03$ &$0.04^{+0.19}_{-0.17}$& $-0.12 \pm 0.08$\\
& $[-0.30, 0.22]$ &$[-0.35, -0.03]$&$[-0.25, 0.37]$& \\
$\acp(B^\pm \to \pi^\pm K^0)$&$+0.18 \pm 0.24$&$-0.21 \pm 0.18 \pm 0.03$& $0.10^{+0.43}_{-0.34}$&
$-0.05 \pm 0.14$ \\
&$[-0.22, 0.56]$& $[-0.51, 0.09]$&$[-0.53, 0.82]$& \\ \hline
\end{tabular}\end{center}
\end{table}

%%%%%%%%%%%%%%%%%%%%%%%%%%%%%%%%%%%%%%%%%%%%%%%%%%%%%%%%%%%%%%%%%%%%%%%%%%%%%%
\newpage

\newpage

\newpage

%%%%%%%%%%%%%%%%%%%%%%%%%%%%%%%%%%%%%%%%%%%%%%%%%%%%%%%%%%%%%%%%%%%%%%%%%%
%%%%%%%%%%%%%%%%%%%%%%%%%%%%%%%%%%%%%%%%%%%%%%%%%%%%%%%%%%%%%%%%%%%%%%%%%%
\newpage

\newpage

\newpage

\newpage

\end{document}